\documentstyle[12pt]{article}

\renewcommand{\thefootnote}{\fnsymbol{footnote}}
\newcommand{\EQ}{\begin{equation}}
\newcommand{\EN}{\end{equation}}
\newcommand{\bea}{\begin{eqnarray}}
\newcommand{\ena}{\end{eqnarray}}
\newcommand{\vs}[1]{\vspace{#1 mm}}

\renewcommand{\a}{\alpha}
\renewcommand{\b}{\beta}

\newcommand{\G}{\Gamma}

\newcommand{\uda}{\nearrow \kern-1em \searrow}
\newcommand{\half}{{1 \over2}}

\newcommand{\cF}{{\cal{F}}(a)}
\newcommand{\La}{\Lambda}
\newcommand{\cFo}{{\cal{F}}_1(a)}
\newcommand{\la}{\lambda}
\newcommand{\nc}{{N_c}}
\newcommand{\halfn}{(\half)_n}
\makeatletter
\def\eqnarray{%
 \stepcounter{equation}%
 \let\@currentlabel=\theequation
 \global\@eqnswtrue
 \global\@eqcnt\z@
 \tabskip\@centering
 \let\\=\@eqncr
 $$\halign to \displaywidth\bgroup\@eqnsel\hskip\@centering
 $\displaystyle\tabskip\z@{##}$&\global\@eqcnt\@ne
 \hfil$\displaystyle{{}##{}}$\hfil
 &\global\@eqcnt\tw@$\displaystyle\tabskip\z@{##}$\hfil
 \tabskip\@centering&\llap{##}\tabskip\z@\cr}
\makeatother

\begin{document}

\oddsidemargin 5mm

\begin{titlepage}
\setcounter{page}{0}
\begin{flushright}
EPHOU 96-006\\
September 1996
\end{flushright}

\vs{6}
\begin{center}
{\Large  Prepotential of $N=2$ Supersymmetric  Yang-Mills Theories \\
in the Weak Coupling Region}

\vs{6}
{\large
Takahiro
Masuda
\footnote{e-mail address: masuda@phys.hokudai.ac.jp}
\\ and \\
Hisao Suzuki\footnote{e-mail address: hsuzuki@phys.hokudai.ac.jp}}\\
\vs{6}
{\em Department of Physics, \\
Hokkaido
University \\  Sapporo, Hokkaido 060 Japan} \\
\end{center}
\vs{6}

\centerline{{\bf{Abstract}}}

We show how to obtain the explicite form of the low energy quantum 
effective action for $N=2$ supersymmetric Yang-Mills theory in the 
weak coupling region from the underlying hyperelliptic Riemann surface. 
This is achieved by evaluating the integral representation of the fields 
explicitly.  We calculate the leading instanton corrections for 
the group $SU(\nc), SO(N)$ and $SP(2N)$ and find that the one-instanton 
contribution of the prepotentials for the these group coincide with 
the one obtained recently by using the direct instanton caluculation.  
\end{titlepage}
\newpage

\renewcommand{\thefootnote}{\arabic{footnote}}
\setcounter{footnote}{0}

 There has been much pregress in the study of N=2 four dimensional 
supersymmetric gauge theories. Originated from the work of Seiberg and 
Witten for the analysis of $SU(2)$ Yang-Mills theories\cite{SW}, the 
framework has been extended to higher rank gauge groups with or without 
matter hypermultiplets\cite{KLTY,APS,HO,Hanany,DS,BL,AS,MW}. Although the 
basic framework has been studied extensively so far, the explicit 
evaluation of the prepotential is a difficult subject. Up to now, 
determination of the prepotential has been  achieved mainly by solving 
Picard-Fuchs equations\cite{CDF,KLT}. Although  more direct method is 
known to be available for $SU(2)$ gauge groups\cite{Matone}, solving 
Picard-Fuchs equation seems to be the best  tool for the investigation 
when these equations turns out to be solved by the well known functions. 
The  $N=2$ $SU(2)$ supersymmetric Yang-Mills theories with massless 
hypermultiplets\cite{IY} and  $SU(3)$ super Yang-Mills theory without 
matters\cite{KLT} have been analyzed elegantly
 by using the method, whereas the method given in Ref.\cite{Matone} is 
useful only for the group $SU(2)$. 
However, the situation changes drastically when we find that Picard Fuchs 
equation cannot be solved by any special functions. This situation generally 
occures for the theories with massive hypermultiplets or  higher rank gauge 
groups. In these cases, we  expect that  the explicit evaluation of the 
integral representation of the solutions is more powerful  than finding the 
solutions of the differential equations. As a matter of fact,  the analytic 
properties of the functions are studied mostly by using the integral 
representations of the functions rather than the differential equations 
themselves even in the theory of special functions.  

In our previous paper, we have evaluated the prepotential of $SU(2)$ 
Yang-Mills theories with massive hypermultiplets by  analyzing the integral 
representations explicitly\cite{MS}, and discussed how these are related 
to the results for the massless cases which have been obtained by using 
Picard Fuchs equations\cite{IY}. 
Therefore, it is natural to expect that this approach is also powerful 
even for the theores with higher rank gauge groups. 

In this paper, we show how to evaluate the prepotential for the gauge 
groups $SU(\nc)$, $SO(N)$ and $SP(2N)$ without matter hypermultiplets in 
weak coupling region. 
Since we cannot rely on the analysis of any special functions, the 
corresponding method for the analysis will be quite different from our 
previous analysis applied for the group $SU(2)$\cite{MS}.  We just use the  
technique used in our previous paper\cite{MS} only for showing the 
equivalence of our expression and the results for $SU(3)$ Yang-Mills 
theory\cite{KLT} where the solutions are  written by Appell hypergeometric 
functions of type $F_4$.
It turns out that the one instanton contribution of the prepotentials for 
the groups $G=SU(N), SO(N)$ and $SP(2N)$ agree completely with the one 
obtained recently by using the direct instanton calculation\cite{IS}.  

In writting this paper, we recieved a preprint \cite{DKP}
 which contains some of our results for $SU(N_c)$.

We will first consider $N=2$ supersymmetric $SU(N_c)$ gauge theories 
without matter hypermultiplets\cite{KLT} in detail. The theory has an 
$N_c -1$ complex dimensional moduli space of vacua which are parameterized 
by the expectation value of the higgs fields as
\bea
<\phi> = \sum_{i=1}^{N_c} e_i H_i = diag[a_1,...,a_{N_c}], 
\ena
where $H_i$ are the generators of the Cartan sub-algebra of $U(N_c)$ and
\bea
\sum_{i=1}^{N_c} a_i =0.
\ena
The  fields $a_D^i$ dual to $a_i$ can be defined as
\bea
a_D^i = {\partial {\cal{F}}(a) \over \partial a_i},
\ena
where ${\cal{F}}(a)$ is the prepotential.
The curve describing the space of vacua can be identified as
\bea
y^2= \prod_{k=1}^{N_c}(x-e_k)^2 - \Lambda^{2N_c},
\ena
where $e_i$ is the value of the classical moduli space with a constraint;
\bea
 \sum_{i=1}^{N_c} e_i =0.
\ena
It should be noted that the classical root of $y$, $e_i$ splits into 
$e_i^+$ and $e_i^-$ in $(4)$.

The meromorphic one form $\lambda$ can be defined as
\bea
\la={dx \over 2\pi i} {x \over y} [\prod_{k=1}^{N_c}(x-e_k)]'.
\ena
The dual pair of fields 
$a$ and $a_D$ can be written as periods of a meromorphic one form 
$\lambda$ on the curve as
\bea
a_i = \oint_{\a_i} \lambda, \qquad a_D^i = \oint_{\b_i} \lambda,
\ena
where $\a^i, \b_i$ form a basis of homology cycles on the curve.
Therefore, we can compute the prepotential ${\cal{F}}(a)$ once we can 
evaluate $a_i$ and $a_D^i$ as functions of $e_k$ and $\Lambda$.

We are going to evaluate $a^i$ in the weak coupling region. 
To begin with, we expand the meromorphic defferential with respect to 
$\Lambda^{2N_c}$
by using an expansion
\bea
{1\over y} = \sum_{n=0}^\infty { (\half)_n \Lambda^{2N_cn} 
\over n! (x-e_1)^{2n+1}\cdot\cdot\cdot(x-e_{\nc})^{2n+1}},
\ena
where $(a)_n$ is defined as
\bea
(a)_n = {\G(a+n)\over \G(a)}.
\ena
After the use of a partial integration, we have
\bea
a_i = \sum_{k=1}^{\nc} \oint_{\a^i} {dx \over 2\pi i} {x \over x-e_k} 
+ \sum_{n=1}^\infty \oint_{\a^i}{dx \over 2 \pi i}{ \halfn \La^{2\nc n} 
\over n! 2n (x-e_1)^{2n}\cdot\cdot\cdot(x-e_{\nc})^{2n}}.
\ena
Originally, the $\a_i$ circle was chosen enclosing the roots of $y$ $e_i^+, 
e_i^-$. In our expression both of them shrink to the classical value $e_i$. 
Therefore, we can take the contour enclosing $e_i$ as $\a^i$ cycle to obtain
\bea
a_i &=& e_i + \sum_{n=1}^\infty {\halfn \La^{2\nc n} \over n! 2n}
\sum_{\scriptstyle m_1,...m_{\nc},m_i=0\atop\scriptstyle m_1+
\cdot\cdot\cdot m_{\nc} = 2n-1} \prod_{\scriptstyle k\atop\scriptstyle 
k\neq i} {\G(2n+m_k)\over \G(2n)\G(m_k+1)}(e_k-e_i)^{-2n-m_k} \nonumber\\
&=& e_i +  \sum_{n=1}^\infty {\halfn \La^{2\nc n} \over n! (2n)!}
({\partial\over \partial e_i})^{2n-1}\prod_{k,k\neq i}(e_k-e_i)^{-2n}.
\ena
In the case of the dual fields $a_D^i$, we have to use analytic 
continuation because of the logarithmic singularity. 
For such purpose, we write the meromorphic one form $\la$ by 
Barnes-type representation as
\bea
\la = {dx \over 2\pi i} \int_{-i\infty}^{i\infty}{ds \over 2\pi i}
{\G(-s)\G(s+1/2) \over \G(1/2)2s}\prod_{k=1}^\nc (x-e_k)^{-2s}(-\La^{2\nc})^s,
\ena
where the path of integration is taken around the poles at $s=0,1,2,...$.
This expression can be obtained by considering Barnes-type 
representation for $(8)$ and use a partial integration. Note also that 
we can obtain the strong coupling expansion of $\la$ by taking the poles 
at $s=-1/2,-3/2,...$  

The $\b_{ij} = b_i - b_j$ cycle consists of the circle enclosing the 
root $e_j^+$ and $e_i^+$, which can be written as two times the line 
integral of $\la$ from $e_j^+$ to $e_i^+$. When we use the expression 
$(12)$, these roots shrink to the classical value. If we replace the 
contour integral to the line integral from $e_j$ to $e_i$, we have to 
subtract the contribution from the circle around  $e_i^-$ and $e_j^-$, 
which can be evaluated  as the half of the $\a^i$ and $\a^j$. 
We therefore obtain an expression of $a_D$ as follows; 
\bea
a_D^{ij} \equiv a_D^i -a_D^j = 2 \int_{e_j}^{e_i} \la -{1\over2}(a_i-a_j).
\ena
The $a_D^i$ can be obtained from the precedure
\bea
a_D^i = {1\over \nc} \sum_{j=1}^{\nc} a_D^{ij}.
\ena
Although the method of the expansion  for $a$  in $(11)$ is quite 
different from those of $a_D$ $(13)$, these will provide a consistent 
result. As a matter of fact, we can write $a$ by using Barnes-type 
integral representation as
\bea
 a_{ij} &\equiv& a_i -a_j \nonumber\\
&=&  \int_{e_j}^{e_i}{dx \over 2\pi i} \int_{-i\infty}^{i\infty}
{ds \over 2\pi i}{\sin 2\pi s \over \pi}{\G(-s)\G(s+1/2) \over \G(1/2)2s}
\prod_{k=1}^\nc (x-e_k)^{-2s}(-\La^{2\nc})^s,
\ena
which we can derive  by considering the integral ecclosing $e_i$ and 
$e_j$ and by replacing it by line integral. The equivalence of $(11)$ 
and $(15)$ can also be shown by evaluating the poles of $s$ in $(15)$ 
explicitly.

As a check of the expression, let us consider the group  $SU(2)$\cite{SW} 
and $SU(3)$\cite{KLT} where the known expression are written in terms of 
the symmetric polynomial of roots rather than roots themselves\cite{KLT}.
In the case of $ SU(2)$, the expression $(11)$ leads to
\bea
a_1 = -a_2 = e_i + \sum_{n=1}^{\infty}{\halfn \La^{2\nc n} \over n! 2n}
{\G(4n-1)\over \G(2n)^2}(e_2-e_1)^{-4n+1}.
\ena
By using Legendre's duplication formula
\bea
\G(2z) = 2^{2z-1}\pi^{-1/2}\G(z)\G(z+1/2),
\ena
we find that $(16)$ can be written as
\bea
a_1 = -a_2 = \sqrt{u} F(-1/4,1/4;1;{\La^4 \over u^2}),
\ena
where we have set $e_1 = -e_2 = \sqrt{u}$. This expression is identical to 
the result obtained in Refs.\cite{SW,CDF,KLT} 
Quite similarly, we can find that $a_D^{12}$ coincides with the known 
expression.

Rather non-trivial check of our expression is for $G=SU(3)$\cite{KLT} 
where the known expression is written in terms of Appell functions of 
type $F_4$ with respect to the variables $u$ and $v$, which are defined by
\bea
(x-e_1)(x-e_2)(x-e_3) = x^3 - ux -v.
\ena
Let us first consider $a_{23}$. From the equation $(15)$, we have
\bea
a_{23} &=& \int_{-i\infty}^{i\infty} {ds \over 2\pi i}{\sin2\pi s\over \pi}
{\G(-s)\G(s+1/2)\over 2s \G(1/2)}{\G(-2s+1)^2\over\G(-4s+2)}
(-\Lambda^6)^s\nonumber\\
&{}&\qquad \times (e_2-e_3)^{-4s+1}(e_3-e_1)^{-2s} 
F(2s,-2s+1;-4s+2;{e_3-e_2 \over e_3-e_1}).
\ena
These are written by root variables. We are going to re-write 
this expression in terms of symmetric polynomial  which is $u$ 
and $v$ in $(19)$. For this purpose, we are going to apply a method 
used in the case of $SU(2)$ super-Yang Mills theories with massive 
hypermultiplets\cite{MS}. Before doing so, we have to choose the branch 
for the low energy expression. We will choose the branch that 
$\vert {e_3-e_2 \over e_3-e_1}\vert$ is in the neiborfood of zero. 
By using  the following quadratic transformation\cite{HTF};
\bea
F(a,b;2b;z) = (1-z)^{-{1\over2}a}F(\half a, b - \half a;b+ \half;
{z^2 \over 4(z-1)}),
\ena
 for $a=2s, b=-2s+1$, we get
\bea
a_{23} &=& \int {ds \over 2\pi i}{\G(-s) 2^{2s-1} \pi^\half \over 
\G(s+1) \G(-2s+{3\over2})}(e_2-e_3)^{-4s+1}(e_3-e_1)^{-s}(e_2-e_1)^{-s}
\nonumber\\
&{}&\qquad \times F(s,-3s+1;-2s+{3\over2};{(e_2-e_3)^2 \over 4(e_3-e_1)
(e_1-e_2)}),
\ena
where we have also used various identities of gamma functions.

This expression is symmetric with respect to $e_1$ and $e_2$. In order to 
obtain fully symmetric form, we use the following cubic transformation
\cite{HTF};
\bea
F({1\over3}-a,3a;2a+{5\over6};z) = (1-4z)^{-3a}F(a,a+{1\over3};2a+{5\over6};
{27z\over (4z-1)^3}),
\ena
for $a=-s+1/3$. By this procedure, we obtain
\bea
a_{23} &=& -\int_{-i\infty}^{i\infty} {ds \over 2\pi i}{\G(-s) 2^{2s-1} 
\pi^\half \over \G(s+1) \G(-2s+{3\over2})}\Delta^{-2s+\half}D^{3s-1}\nonumber\\
&{}&\qquad \times F({1\over3}-s,{2\over3}-s;-2s+{3\over2};{27\Delta
\over 4 D^3}),
\ena
where
\bea
\Delta &=& (e_1-e_2)^2(e_2-e_3)^2(e_3-e_1)^2 = 4u^3-27v^2, \nonumber\\
D &=& {1\over2}[(e_1-e_2)^2+(e_2-e_3)^2+(e_3-e_1)^2]=3u..
\ena
It should be noted that these are written in the form which is totally 
symmetric  with respect to roots variables so that these can be expressed 
in terms of $u$ and $v$.

In order to write this expression in terms of Appells functions, we make 
use of the formula\cite{HTF};
\bea
F(a,b;c;z)&=&{\G(c)\G(c-a-b)\over \G(c-a)\G(c-b)}F(a,b;a+b-c+1;1-z)\nonumber\\
&+&{\G(c)\G(a+b-c) \over \G(a)\G(b)}(1-z)^{c-a-b}F(c-a,c-b,c-a-b+1;1-z),
\ena
 apply a formula\cite{HTF}
\bea
F(a,b;c;z) =(1-z)^{c-a-b}F(c-a,c-b;c;z),
\ena
then we finally find that the function $a_{23}$ can be written as
\bea
a_{23} = u^\half [ -F_4({1\over6},-{1\over6};{1\over2};1;x;y) + 
({x\over3})^\half F_4({1 \over 3},{2\over3};{3\over2};1;x;y)],
\ena
where the function $F_4$ is the Appel function\cite{HTF} and 
$x={27 \over 4}{v^2\over u^3}$,$y={27 \over 4}{\Lambda^6 \over u^3}$.
 In the branch that  $\vert {e_3-e_2 \over e_3-e_1}\vert$ is in the 
neiborfood of zero, we should apply analytic continuation of 
the variables in $a_{12}$ and $a_{31}$ before usiong the  quartic and 
cubic transformations. The result $a_{12}$ is given by
\bea
a_{12} = 2 u^\half F_4({1\over6},-{1\over6};{1\over2};1;x;y).
\ena
  You can find that the same procedure can be applied for $a_D^{ij}$.
The resulting expressions completely agree with the ones obtained in ref.
\cite{KLT}.

In the case of $\nc >3$, we do not know how to express our result in 
terms of symmetric polynomials with respect to roots $e_k$  because not 
so many  quartic transformations have been known for the hypergeometric 
functions of several variables. Although  the corresponding transformation, 
if it exists, may be powerful when we consider the theories with matter 
multiplets as in the case of $SU(2)$ gauge theories\cite{MS}, we here  
evaluate the integral for $a_D^i$ directly rather than trying to 
represent them in terms of symmetric polynomial. This approach seems to 
be combenient for obtaining the prepotential. Off course, the most 
difficult question is how to get an analytic continuation which is 
consistent with the integrability of the prepotential in $(3)$. In this 
paper, we assume that the integration over $s$ in the expression $(12)$ 
will regularize the function correctly. At present, we cannot justify 
the procedure, because many summations over some other integers appear in 
the course of the evaluation. However, we will show that we can obtain 
the result consistent with the integrability at least up to the leading 
order of the instanton corrections. 

In the expression for $a_D^{ij}$, the singularities appear as double poles 
for the integral with respect to $s$, which  consists of the contribution 
both from $a_i$ and $a_j$. Since $a_i$ and $a_j$ have different 
singularities, it seems not easy to extract these double poles in a consice 
manner. We therefore consider the path from zero to $e_i$ and define
\newcommand{\ta}{{\tilde{a}}}
\bea
\ta_D^i \equiv {1 \over \pi i} \int_0^{e_i} \la-{1\over2}a_i,
\ena
where we have again subtracted the contribution caused by the degeneracy 
of the roots in the expression of $\la$ in $(12)$.
In the weak coupling region, $a_D^{ij}$ can be written as
\bea
a_D^{ij} = \ta_D^i -\ta_D^j.
\ena
From the relation $(14)$, we have
\bea
a_D^i = \ta_D^i - {1\over \nc}\sum_{k=1}^{\nc}\ta_D^k.
\ena
Let us evaluate $\ta_D^i$.   
By parameterizing $x=e_i(1-t)$,
we have
\bea
\ta_D^i = &{}&{1\over\pi i\G(1/2)} \int_{-i\infty}^{i\infty}{ds\over2\pi i}
{\G(-s)\G(s+1/2)\over 2s} \sum_{m_k,k\neq i}\int_0^1  t^{-2s+1+\sum m_k}dt 
\nonumber\\
&\times&\prod_{k, k\neq i}{\G(2s+m_k)\over\G(2s)\G(m_k+1)}
(e_i-e_k)^{-2s+m_k}e_i^{-2s+1+\sum m_k}(-\La^{2\nc})^s -{1\over2} a_i,
\nonumber\\
=  &{}&{1\over\pi i\G(1/2)} \int_{-i\infty}^{i\infty}{ds\over2\pi i}
{\G(-s)\G(s+1/2)\over 2s} \sum_{m_k,k\neq i} {1\over -2s+\sum m +1}
\nonumber\\
&\times& \prod_{k, k\neq i}{\G(2s+m_k)\over\G(2s)\G(m_k+1)}(e_i-e_k)^{-2s+m_k}
e_i^{-2s+1+\sum m_k}(-\La^{2\nc})^s-{1\over2}a_i. 
\ena
Strictly speaking, in order to performing the $t$-integral, we have to 
make use of the analytic continuation from the strong coupling expression 
which takes the pole $s=-1/2,-3/2,...,$ and after performing integral we 
have to use the analytic continuation to the weak coupling region.
Let us evaluate the poles separately. At first, we are going to evaluate 
the pole at $s=0$. The double pole arises in the case of  $m_k=0$ for 
all $k$, and single poles appear when $m_k \neq 0$ for one of $k$. We 
can evaluate the pole directly to find that the contribution from the 
zeroth order is given by
\bea
\ta_D^{i(0)} = &{1 \over 2\pi i}&\lbrace -\sum_{k=1}^{\nc}(e_i-e_k)
\ln{(e_i-e_k)^2\over \La^2} + [2+(\psi(1/2)-\psi(1))/\nc]\sum_k (e_i-e_k) 
\nonumber\\
&-&\sum_{k=1}^{\nc} e_k \ln{e_k^2\over \La^2}\rbrace.
\ena
From the relation $(30)$ we have
\bea
a_D^{i(0)} = &{1 \over 2\pi i}&\lbrace-\sum_{k=1}^{\nc}(e_i-e_k)
\ln{(e_i-e_k)^2\over \La^2}\nonumber\\
 &+& [2+(\psi(1/2)-\psi(1))/\nc]\sum_k (e_i-e_k) \rbrace,
\ena
which has exactly required form of $a_D^i$ in the weak coupling 
region\cite{KLTY}.
We next evaluate the contribution from $s=1$. The double poles appear in 
the case $\sum m_k =1$ in $(33)$. Other terms have single poles so that 
we can evaluate them without any analytic continuation by going back to 
the original expression for $\la$ for $s=1$. We can use the following 
decomposition
\bea
{1 \over \prod_{k=1}^{\nc}(x-e_k)^2} = \sum_{k=1}^{\nc}\lbrace 
{1 \over \prod_{l\neq k}(e_k-e_l)^2}{1\over (x-e_k)^2}+
{\partial \over \partial e_k}[{1 \over \prod_{l\neq k}(e_k-e_l)^2}]
{1\over (x-e_k)}\rbrace.
\ena
Of all the expansion of $(30)$ by using the expansion $(36)$, the term 
containing $1/(x-e_i)^2$ can be evaluated from the expression $(33)$ 
as the term satisfying $\sum m_k =0$ and the term of the form $1/(x-e_i)$ 
was evaluated as double pole in $(33)$ so that we should subtract it. 
Other terms can be integrated explicitly. The result up to the leading 
order turns out to be
\bea
a_D^i =&{}& {i \over 2\pi}\sum_k (a_i-a_k)\ln{(e_i-e_k)^2\over\La^2} - 
{i\over \pi}\sum_k(e_i-e_k)-{i\over2\pi}[{(\psi(1/2)-\psi(1))/ \nc}]
\sum_k(a_i-a_k)\nonumber\\
& &-{i\La^{2\nc} \over 8\pi}{\partial \over \partial e_i}
[\sum_k {1 \over \prod_{l\neq k}(e_k-e_l)^2}],
\ena
which can be written by $a$ in the following form;
\bea
a_D^i =&{}& {i \over 2\pi}\sum_k (a_i-a_k)\ln{(a_i-a_k)^2\over\La^2}- 
{i\over2\pi}[2+{(\psi(1/2)-\psi(1))/\nc}]\sum_k(a_i-a_k)\nonumber\\
& &-{i\La^{2\nc} \over 8\pi}{\partial \over \partial a_i}[\sum_{k=1}^{\nc} 
{1 \over \prod_{l\neq k}(a_k-a_l)^2}].
\ena
The prepotential at this order can be obtained as
\bea
\cF={i\over 4\pi}\sum_{i<j}(a_i-a_j)^2\ln{(a_i-a_j)^2/\La^2}+
{\tau_0\over 2\nc}\sum_{i<j}(a_i-a_j)^2 + \cFo,
\ena
where $\tau_0$ is the bare coupling;
\bea
\tau_0 = {i\over 2\pi}(2\ln 2-3\nc),
\ena
and the one-instanton contribution $\cFo$ is given by
\bea
\cFo = -{i\La^{2\nc}\over 8\pi}\sum_{k=1}^{\nc} {1 \over \prod_{l\neq k}
(a_k-a_l)^2}
\ena
Therefore, our method for analytic continuation is consistent 
with the integrability of the prepotential at least up to the 
leading order of the instanton corrections. It should be noted that the 
expression (39)  agrees completely with the known results for 
$G=SU(2)$ and $SU(3)$\cite{KLT}\footnote{The bare coupling agrees 
when we correct a misprint in ref.\cite{KLT}.}. 
Moreover, it coincides with the result obtained by the direct 
instanton method for $SU(\nc)$\cite{IS}.

Let us consider the theories with other gauge groups. It is straightforward 
to apply the method to other groups. We are now going to list the curve 
and the one instanton contribution of other classical groups.

For $SO(2N+1)$, the curve is identified as\cite{DS}
\bea
y^2 = \prod_{k=1}^{N}(x^2-e_k^2)^2 - \La^{2(2N-1)}x^2.
\ena
From this curve, we can calculate the one instanton contribution as
\bea
\cFo = - {i \La^{2(2N-1)} \over 32 \pi} \sum_{k=1}^{N}
{1 \over \prod_{l \neq k} (a_k^2 -a_l^2)^2}.
\ena

For $SO(2N)$, the curve is given by\cite{BL}
\bea
y^2 = \prod_{k=1}^{N}(x^2-e_k^2)^2 - \La^{2(2N-1)}x^4,
\ena
from which the one instanton contrubution is obtained as
\bea
\cFo = - {i \La^{4(N-1)} \over 32 \pi} \sum_{k=1}^{N}{ a_k^2 \over 
\prod_{l \neq k} (a_k^2 -a_l^2)^2}.
\ena
In the case of $SO(4)$, you can find the decomposition to $SU(2)\times SU(2)$.

The Weierstrass form of the curve for the groups $SP(2N)$ is given by
\bea
y^2 = P^2(x)  - \La^{2(N+1)} P(x).
\ena
where
\bea
P(x) = x^2\prod_{k=1}^N (x^2 - e_k^2).
\ena
The equivalent  Riemann surface is\cite{MW} 
\bea
f= (z+{\La^{2(N+1)} \over z})^2 - 4P(x)=0,
\ena
whose equivalence can be checked by evaluating the periods.
The meromorphic one form of the curve is obtained as
\bea
\la = {dx \over 2\pi i} \int_{-i\infty}^{i\infty}{ds \over 2\pi i}
{\G(-s)\G(s+1/2) \over \G(1/2)s}x^{-2s}\prod_{k=1}^N (x^2-e_k^2)^{-s}
(-\La^{2(N+1)})^s,
\ena
where the integration over $s$ takes the poles at $s=0,1,...\infty$.
It can be shown that the classical part of the prepotential agrees 
with the general form and we find that the one instanton contrubution 
is given by
\bea
\cFo = - {i \La^{2(N+1)} \over 4 \pi} {(-1)^N \over \prod_{k=1}^N a_k^2}.
\ena
Note that all these results agree completely with the one obtained 
by the direct instanton method\cite{IS}.

We have shown how to calculate the effective action of $N=2$ 
supersymmetric Yang-Mills theories without matter hypermultiplets. 
 At this moment, it is not clear whether  our method of analytic 
continuation is consistent  with the integrability of the prepotential 
at all orders. Although we could obtain rather compact expression 
for $a$, we have not been able to obtain the general form of the dual 
fields $a_D$. Actually the calucuation of the next leading order seems 
very complicated and  requires more simplification of our method.

When we consider the $SU(2)$ theories with matter hypermultiplets, 
the use of symmetric polynomial has been shown to be useful\cite{IY,MS}. 
Therefore, it is natural that we can obtain a compact expression 
by using the symmetric polynomial of the root variables even for the 
theories having higher rank gauge groups.  It seems interesting to 
analyze the quartic and quadratic transformations in these cases.  

We would like to thank K. Suehiro for discussions.

\end{document}